\documentclass[aps,prb,twocolumn,superscriptaddress,showpacs]{revtex4}

\usepackage{graphicx, latexsym, verbatim}
\usepackage{amssymb, amsmath}
\usepackage[ansinew]{inputenc}

\newcommand{\ud}{\mathrm{d}}

\newcommand{\Hb}{\mathbf{H}}
\newcommand{\Gb}{\mathbf{G}}
\newcommand{\Vb}{\mathbf{V}}

\newcommand{\Sb}{\mathbf{S}}
\newcommand{\Xb}{\mathbf{X}}

\newcommand{\bEqa}{\begin{eqnarray}}
\newcommand{\eEqa}{\end{eqnarray}}
\newcommand{\Tr}{{\rm Tr}}

\begin{document}


\title{Electronic transport in Si nanowires: Role of bulk and surface disorder}



\author{Troels Markussen}
\affiliation{MIC - Institute for Micro- and Nanotechnology, NanoDTU, Technical University of Denmark, DK-2800 Kgs. Lyngby}

\author{Riccardo Rurali}
\affiliation{Departament d'Enginyeria Electrònica, Universitat Autònoma de Barcelona 08193 Bellaterra, Spain}

\author{Mads Brandbyge}
\affiliation{MIC - Institute for Micro- and Nanotechnology, NanoDTU, Technical University of Denmark, DK-2800 Kgs. Lyngby}

\author{Antti-Pekka Jauho}
\affiliation{MIC - Institute for Micro- and Nanotechnology, NanoDTU, Technical University of Denmark, DK-2800 Kgs. Lyngby}


\date{\today}

\begin{abstract}
We calculate the resistance and mean free path in long metallic and semiconducting silicon nanowires (SiNWs) using two different numerical approaches: A real space Kubo method and a recursive Green's function method. We compare the two approaches and find that they are complementary: depending on the situation a preferable method can be identified. Several numerical results are presented to illustrate the relative merits of the two methods.
Our calculations of relaxed atomic structures and their conductance properties are based on density functional theory without introducing adjustable parameters. Two specific models of disorder are considered: Un-passivated, surface reconstructed SiNWs are perturbed by random on-site (Anderson) disorder whereas defects in hydrogen passivated wires are introduced by randomly removed H atoms. The un-passivated wires are very sensitive to disorder in the surface whereas bulk disorder has almost no influence. For the passivated wires, the scattering by the hydrogen vacancies is strongly energy dependent and for relatively long SiNWs ($L>200\,$ nm) the resistance changes from the Ohmic to the localization regime within a 0.1 eV shift of the Fermi energy. This high sensitivity might be used for sensor applications.


\end{abstract}

%


\maketitle

\section{Introduction}
Semiconducting nanowires are very promising building block for future nanoelectronic and nanophotonic applications as witnessed by several recently demonstrated devices\cite{LieberScience2001, GudiksenLieber2002, SamuelsonMatToday, CuiLieberNanoLett2003, SamuelsonPhysicaE, WuLieberNature2004}. Silicon nanowires (SiNWs) are especially attractive candidates due to their compatibility with conventional Si-technology and due to the accurate control of diameter and electronic properties during synthesis\cite{WuLieberNanoLett2004}. Furthermore, in recent years SiNWs have been applied as label-free real-time chemical and biological sensors with very high sensitivity and, e.g., capability of single virus detection \cite{LieberMatToday2005}.

Thin SiNWs with diameters below 5 nm have been synthesized by several groups. Recently, Ma \textit{et al.}\cite{DDDMa} obtained very thin wires grown in the $\langle 1 1 0\rangle$ and $\langle 1 1 2\rangle$ directions with diameters down to $1.3\,$nm, and Holmes \textit{et al.}\cite{HolmesScience2000} have previously reported 4-5 nm $\langle 1 0 0\rangle$ and $\langle 1 1 0\rangle$ SiNWs. Wu \textit{et al.}\cite{WuLieberNanoLett2004} recently demonstrated that the growth directions depend on the diameter, which can be controlled by the size of a catalytic nanoparticle\cite{LieberAPL2001}.

Concerning theoretical modelling of the structural properties, Rurali and Lorente\cite{NicolasPRL} recently showed, using \textit{ab initio} calculations, that thin $\langle 1 0 0\rangle$ un-passivated SiNWs could be either metallic or semi-metallic depending on the surface reconstruction. In another recent work, Singh \textit{et al.} \cite{SinghNanoLett2005} theoretically studied pristine $\langle 110\rangle$ SiNWs and found that these were indirect band gap semiconductors. Very recently, Fernández-Serra \textit{et al.} \cite{BlasePRL} used \textit{ab initio} calculations to study the surface segregation of dopants in both passivated and un-passivated SiNWs.

The large surface to bulk ratio in the thin wires imply that surface effects such as defects, vacancies or adatoms will have a large influence on the transport properties. This was experimentally demonstrated by Cui \textit{et al.}\cite{CuiLieberNanoLett2003} showing increased mobilities after passivation of surface defects. Also, electron-phonon scattering might be suppressed in thin wires. Indeed, recent experiments by Lu \textit{et al.}\cite{LuLieberPNAS2005} indicated ballistic transport in undoped Si/Ge core-shell wires at room temperature with an estimated phonon scattering mean free path (MFP) $l_{ph}>500\,$nm. This might imply that even at room temperature defects could be the most important scattering source, and a thorough understanding of the scattering processes is thus required.

A number of transport calculations have been reported for wires with various diameters. Das \textit{et al.} \cite{Das2005} used the Boltzmann equation to calculate the carrier mobility in relative thick ($d=10-90\,$nm) GaAs wires focusing on the diameter dependence. Sundaram \textit{et al.} \cite{Sundaram2004} also used the Boltzmann equation to study surface effects on the transport in large diameter wires.  Zheng \textit{et al.} \cite{Lake2005} applied a tight-binding model of a hydrogen passivated wire and studied the effect of wire thickness on the band gap, effective masses and transmission.

Real SiNWs with lengths up to the scale of micrometers consist of millions of atoms and are likely to have many randomly placed defects. To our knowledge, no theoretical works concerning SiNWs, based on \textit{ab initio} methods and including many scattering events, have yet been published.

A calculation of the conductance of a SiNW with many randomly positioned defects puts strong requirements on the method. The quasi one-dimensional nature of the SiNWs requires on the one hand an atomistic model taking quantum effects and charge transfer around the defect into account. On the other hand the method should be able to treat more than $10^5$ atoms and include many scattering events due to the $\mu$m length of the wires.

In this work, two methods are used to study the effect of disorder including many randomly placed H-vacancies in H-passivated long SiNWs. Both methods are  based on \textit{ab initio} calculations and scale linearly with the length of the sample.
The first approach uses a relatively new real-space method, developed by S. Roche and D. Mayou and co-workers over the last decade to study transport properties primarily in carbon nanotubes and quasi-periodic systems.\cite{RochePRB99, RocheMayouPRL97, MayouPRL2000, RochePRL2001, TriozonPRB2004, RochePRL2004, RochePRB2005, RochePRL2005, RocheNanoLett2005} The method is based on the Kubo-Greenwood formalism \cite{Kubo1957, Greenwood} rewritten in a real-space framework, and we will refer to it as the Kubo method. The second and more well-known approach is based on the Landauer formula where the conductance is found by recursive calculations of Green's functions (GFs). We will refer to this as the GF method.

The Kubo method was shown to predict the elastic MFP at the Fermi energy in randomly disordered carbon nanotubes (CNTs)\cite{TriozonPRB2004}, in agreement with a Fermi's golden rule estimate\cite{TodorovNature}. Besides that, we are aware of no comparison with other theoretical methods.  One of the primary goals of this paper is to report such a comparison over for several energies. We show that the Kubo and GF methods are in general in qualitative agreement, however, with quantitative differences especially around band edges. We analyze the pros and cons of the two methods and give an assessment of when they should be applied and which quantities they can calculate.

The rest of the paper is organized as follows. In section \ref{Methods} we summarize the two numerical methods and describe how a Hamiltonian for a long SiNW is constructed from \textit{ab initio} calculations. Results concerning both un-passivated as well as hydrogen passivated SiNWs are presented in section \ref{Results}. We end up with a discussion of the applied methods and the results in section \ref{Conclusion}.

\section{Methods} \label{Methods}
\subsection{Real space Kubo method}
In the real space Kubo method, which is derived from the Kubo-Greenwood formula, transport properties are determined by calculating the time propagation of wave packets in real space. The central quantity is the time- and energy dependent diffusion coefficient, $\mathcal{D}(E,t)$, defined by
\begin{equation}
\mathcal{D}(E,t) = \frac{1}{t}\frac{\Tr\{(\Xb(t)-\Xb(0))^2\,\delta(E-\Hb) \} }{\Tr\{\delta(E-\Hb)\}}, \label{Diffusion-coefficient-def}
\end{equation}
where $\Xb(t)$ is the position operator along the wire direction written in the Heisenberg representation, $\Hb$ is the Hamiltonian matrix and the trace $\Tr\{\delta(E-\Hb)\}$ is the total electronic density of states (DOS). Following Triozon \textit{et al.} \cite{TriozonPRB2002, TriozonPRB2004}
an efficient evaluation of the traces can be carried out by using a relative modest number ($<10$) of \textit{random phase states}, $|\psi_r\rangle$. The coefficients of $|\psi_r\rangle$ which are labeled from $-N/2$ to $N/2$, with $N$ being the total number of orbitals in the wire, are initially non-zero only in the central part of the nanowire, i.e.
\begin{equation}
|\psi_{r}(j)\rangle =   \begin{cases}
    \frac{1}{\sqrt{N_r}}\,e^{2i\pi\alpha(j)}, & \text{for }\,-N_r/2\leq j\leq N_r/2, \\
    0 & \text{otherwise}.
  \end{cases}
 \label{random-phase-state}
\end{equation}
where $\alpha(j)$ is a random number in the interval $[0,1[$ and typically $N_r\sim10^{4}$. The number of random phase states needed to accurately estimate the traces is not known a priori and the convergence of the results must be checked.

Time evolution of the random phase states can be efficiently computed by expanding the time evolution operator, $e^{-i\Hb t/\hbar}$ in the orthogonal set of Chebyshev polynomials. 
Each term in the traces in Eq. (\ref{Diffusion-coefficient-def}) is a local density of state which is calculated using a continue fraction technique.
This is the most time consuming part of the calculations and involves $\sim10^3$ operations with the Hamiltonian for the considered systems to resolve the closely lying energy bands. The convergence of the continued fraction scheme must be separately verified.

The elastic MFP, $l_e$, is found from\cite{TriozonPRB2004}
\begin{equation}
l_e(E)=\frac{{\rm max}\{\mathcal{D}(E,t),\,t>0\}}{v(E)}  \label{mean-free-path},
\end{equation}
where $v(E)$ is the group velocity at energy $E$.

Notice that there is no requirement that the Hamiltonian is periodic. Nor are there any leads that connect to a device area. We also stress that the time it takes to calculate the diffusion coefficient at many energies is not much longer than for a single energy. The reason for this is that the random phase states contain all energy components and the time evolution is energy independent. Moreover, the primary numerical task in the continued fraction scheme is a mapping of the original Hamiltonian to a smaller tridiagonal matrix. This mapping is energy independent, and it is therefore relatively fast to compute the MFP for the whole energy spectrum.

\subsection{Recursive Green's function method}
\begin{figure}[!htb]
\begin{center}
    \includegraphics[angle=90, width=0.47\textwidth]{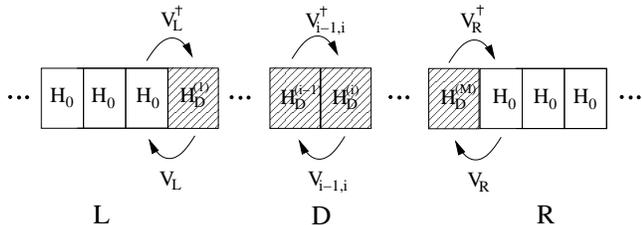}
    \caption{\emph{The device region (D) is divided into M sub-cells. The sub-cells are so large that they only couple to the nearest neighbour cells. Also, the left (L) and right (R) leads consisting of equal unit cells, $\Hb_0$, couple only to the first and last cell of the device region respectively.}}
        \label{Split-device-figure}
\end{center}
\end{figure}

The second numerical method we have applied is based on the Landauer formalism described in detail in e.g. Ref. \onlinecite{Todorov}. The general setup is illustrated in Fig. \ref{Split-device-figure}. A device region (D) is connected to a left (L) and right (R) semi-infinite lead. The device area consists of $M$ sub-cells and is described by the Hamiltonian $\Hb_D$
\begin{equation}
\Hb_D = \left(
\begin{array}{cccc}
  \Hb_D^{(1)}            &    \Vb_{1,2}      &   0    &  \hdots  \\
  \Vb_{1,2}^{\dagger} &    \Hb_D^{(2)}        & \Vb_{2,3} &  \\
  0                & \Vb_{2,3}^{\dagger} &  \ddots& \ddots  \\
   \vdots              &                &  \ddots& \Hb_D^{(M)}
\end{array}
\right). \label{Hamiltonian-matrix}
\end{equation}
The sub-cells are chosen so large that only nearest neighbour cells couple. Generally, the sub-cells do not need to have the same size. The leads described by $\Hb_L$ and $\Hb_R$ respectively are assumed to have a semi-infinite structure consisting of equal unit cells, $\Hb_0$. The coupling matrices between the leads and the device area are denoted $\Vb_L$ and $\Vb_R$. The Hamiltonians are calculated with a non-orthogonal basis set (see section \ref{H-construction}) such that for each Hamiltonian matrix we also have a corresponding overlap matrix $\Sb$.

To calculate the length dependent conductance of the wire, we initially find the surface Green's functions, $\Gb_{L,R}^0(E) =  \left(E\Sb_{L,R}-\Hb_{L,R} \right)^{-1}$, of the isolated leads by a standard decimation procedure.\cite{LoLoRu.84.}  The device area is subsequently grown by adding one sub-cell at a time and calculating the Green's function
\begin{eqnarray}
\Gb_{i}(E) = \left( E\,\Sb_D^{(i)} -\Hb_D^{(i)} - \mathbf{\Sigma}_L(E) - \mathbf{\Sigma}_R(E) \right)^{-1},
\end{eqnarray}
where $\mathbf{\Sigma}_{R}(E)$ describes the coupling to right lead and is defined through $\Gb_{R}^0(E) = \left(E\Sb_0-\Hb_0 - \mathbf{\Sigma}_{R}(E) \right)^{-1}$. $\mathbf{\Sigma}_{L}(E)$ also takes the coupling to the rest of the device area into account and is calculated from the previous Green's function as $\mathbf{\Sigma}_L(E)=\Vb_{i-1,i}^\dagger\Gb_{i-1}(E)\Vb_{i-1,i}$.

In each growth step we calculate the conductance of the wire consisting of the first $i$ sub-cells as
\begin{eqnarray}
g(E,L_i) = \frac{2e^2}{h}\,\Tr\left[\Gb_i^\dagger(E)\,\mathbf{\Gamma}_R(E)\,\Gb_i(E)\,\mathbf{\Gamma}_L(E)\right]  \label{Landauer-conductance-istep},
\end{eqnarray}
where $L_i$ is the length of the grown device region, $\mathbf{\Gamma}_{R,L}(E) = -2\,{\rm Im}\left[\mathbf{\Sigma}_{R,L}(E)\right]$ and where the trace is performed over the states in the device region.

Sample averaging is performed over 200 different configurations giving a mean conductance, $\langle g\rangle$. 
The corresponding resistance is found as $R=1/\langle g\rangle$. For wire length $L\ll\xi$, with $\xi$ being the localization length, the resistance increases linearly as $R(L)=R_0+R_0L/l_e$, defining the MFP, $l_e$.\cite{Datta} In the localization regime the resistance increases exponentially and we calculate the localization length as~\cite{MacKinnon}
\begin{equation}
\xi = -\lim_{L\rightarrow \infty}\frac{2L}{\langle \ln g\rangle},
\end{equation}
where the limit is taken such that $\xi$ is converged.

\subsection{Constructing the Hamiltonian from \textit{ab initio} calculations} \label{H-construction}
The atomic and electronic structure of the SiNWs is found from ab initio calculations using the density functional theory (DFT) package {\sc Siesta}.\cite{siesta-ref} Using first principle calculations it is relatively straightforward to introduce various defects such as vacancies, adatoms or dopants. This is not the case when using standard tight-binding parameters.

We have used a minimal single-$\zeta$ basis set\cite{SiestaBasis}, with 4 orbitals (one 3$s$ and three 3$p$) on the Si atoms and one on the H atoms, to represent the one-electron wave function. The minimal basis set is applied in order to speed up the subsequent transport calculations. We have used norm-conserving pseudopotentials of the Troullier-Martins type~\cite{pseudo} and the Generalized Gradient Approximation~\cite{gga} for the exchange-correlation functional.
The calculations are performed on super-cells containing five wire unit cells (see Fig. \ref{wire-fig}). The reciprocal space has been sampled with a converged grid of $1 \times 1 \times 2$ $k$ points following the Monkhorst-Pack scheme.\cite{monk-pack}

When modelling a defect this is placed in the middle of the supercell to ensure that the effect of the defect is confined to the central region and is not affected by the periodicity and the inter-cell coupling terms $V_{i-1}$ in Eq.(4) are independent on the specific type of defect. The atomic postions of all the atoms in the cell containing the defect and in the two neighboring cells have been fully relaxed, until the maximum force was smaller than 0.04~eV/\AA. This is an important point, because the local distortion induced by the defect can have important consequences on its scattering properties.

To model a long ($L>100\,$nm) SiNW with a random distribution of defects (section \ref{squarewire}) we perform a full \textsc{Siesta} calculation for a pristine wire and for each type of defect.
The wire is grown by adding  pieces from the different calculations ending up with a wire structure as in Fig. \ref{Split-device-figure} and a Hamiltonian as Eq. (\ref{Hamiltonian-matrix}). The sub-cells $\Hb_D^{(i)}$ can either be pristine parts or contain a defect. 

Usually when joining pieces from different calculations the Fermi energies should be aligned to ensure charge conservation. However, for the SiNWs with hydrogen vacancies, a dangling bond (DB) state forms in the band gap and pins the Fermi energy causing a large shift compared to a pristine wire. Alignment of the Fermi energies would therefore lead to an unphysical shift in energies also for sites far away from the vacancy. We therefore align the Si $3s$ on-site energy instead, taking as reference the first Si atom in the supercell.

The minimal basis set implies that we can not expect the conduction bands to be accurately described, and we will thus only focus on the valence bands. Furthermore, the Hamiltonian is truncated by removing the smallest elements, while keeping the band structure constant with a tolerance of $10^{-2}$ eV. This leads to a reduction of non-zero matrix elements of around $80\%$, and we thus effectively end up with a tight-binding-like Hamiltonian.

{\sc Siesta} uses a non-orthogonal atomic basis set giving rise to an overlap matrix $\Sb$. The Kubo method, however, requires an orthogonal basis set and the \textsc{Siesta} output cannot therefore be used directly for which reason a Löwdin transformation is performed. For each \textsc{Siesta} calculation the Hamiltonian, $\Hb$, containing five unit cells described in the non-orthogonal atomic basis, is mapped to a new one, $\widetilde{\Hb}$, with an orthogonal basis:
\begin{equation}
\Hb\rightarrow\widetilde{\Hb}=\Sb^{-1/2}\,\Hb\,\Sb^{-1/2} \label{H-mapping}.
\end{equation}
In the same way as described above, a long wire is built by extracting parts from both pristine and defected orthogonalized Hamiltonians and joining the pieces. The Löwdin transformation leads to a longer range of the basis orbitals and thus more non-zero elements in the truncated Hamiltonian. This in turn implies longer calculation times.

\subsection{Fermi's golden rule}
We next compare the two numerical approaches with results obtained using Fermi's golden rule (FGR). We consider scattering due to the scattering potential $V$ between unperturbed Bloch states, $|n,k\rangle$, of the pristine wire. The transport relaxation rate, $1/\tau_n$, from band $n$ at energy $E$ is found as
\begin{eqnarray}
\frac{1}{\tau_n(E_n(k))} &=& \frac{2\pi}{\hbar}\int\,\ud k'\sum_{m}\overline{|\langle\,k',m\,|V|\,n,k\rangle|^2}\,\nonumber \\ &\quad&\times(1-\cos\theta_{kk'})\,\delta(E_m(k')-E_n(k)) \nonumber \\
&=& \frac{4\pi}{\hbar}\sum_{m}\overline{|\langle\,-k'',m\,|V|\,n,k\rangle|^2}\, n_{\mbox{\tiny{$m$}}}(E) \label{tau-fgr},
\end{eqnarray}
where the summation is over bands, $n_{\mbox{\tiny{$m$}}}(E)$ is the DOS corresponding to band $m$ and the energy of the final state fulfills $E_m(-k'')=E_n(k)=E$. $\theta_{kk'}$ is the angle between the initial and final wavevector. Only scattering from a forward to a backward propagating state contributes to the rate, i.e. $\theta_{kk'}=-\pi$ yielding a factor 2. The perturbing potential, $V$, is for the Anderson on-site disorder (section \ref{metallicResults}) simply a diagonal matrix with the diagonal elements being random numbers in a given interval (see below), and the bar denotes an average over different configurations of $V$. The MFP for electrons in band $n$ is calculated as
\begin{eqnarray}
l_n(E) = v_n(E)\,\tau_n(E) \label{le-fgr-band-n},
\end{eqnarray}
where $v_n(E)$ is the group velocity at energy $E$ in the $n$'th band. We calculate the total MFP as the mean value from the individual contributions from the bands:
\begin{eqnarray}
l_e(E) = \frac{1}{N(E)}\sum_n l_n(E) ,
\end{eqnarray}
where $N(E)$ is the number of bands at energy $E$.


\section{Results}
\label{Results}
\subsection{Surface reconstructed wires} \label{metallicResults}
\begin{figure}[htb]
\includegraphics[width=.3\textwidth]{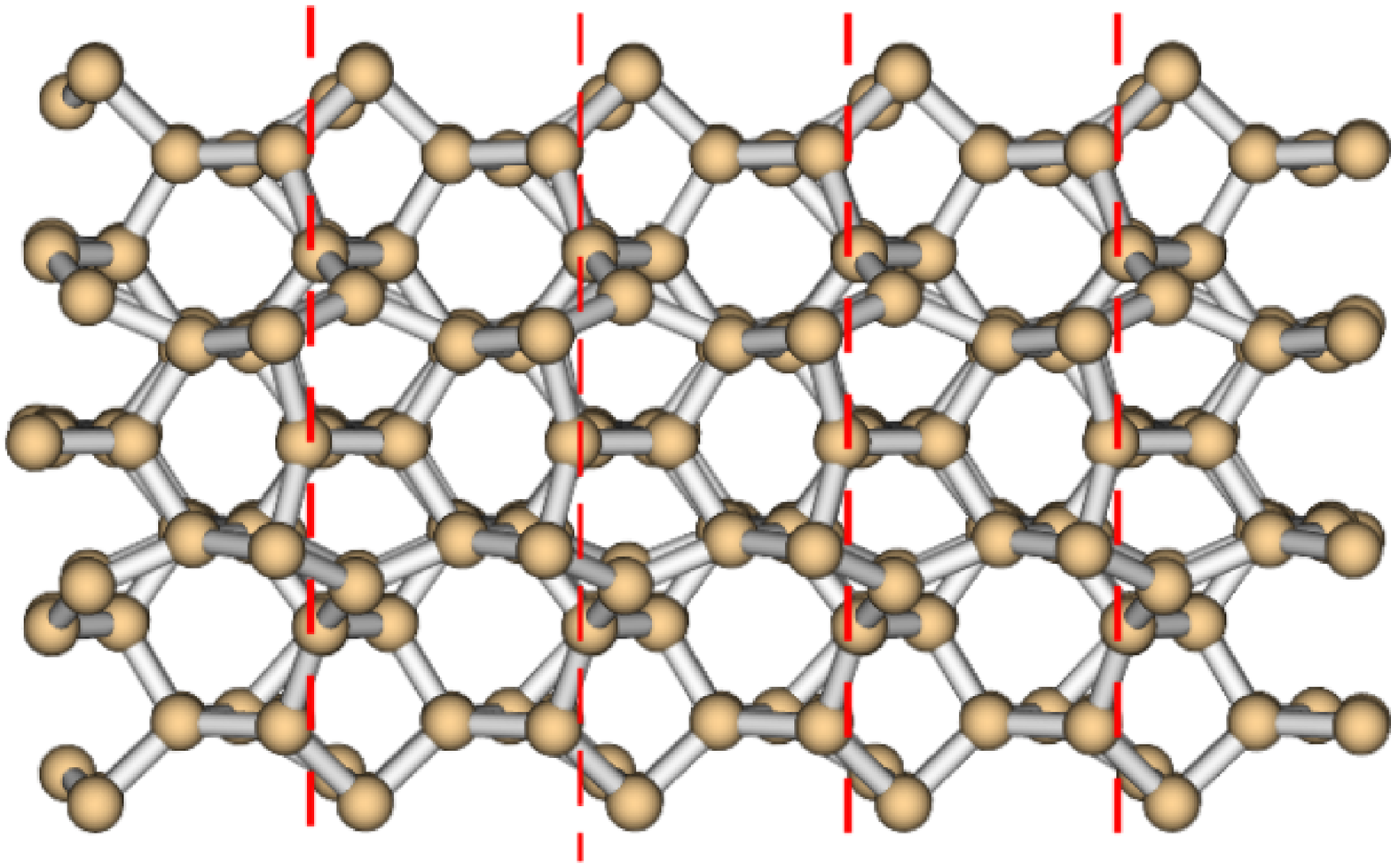}
\includegraphics[width=.2\textwidth]{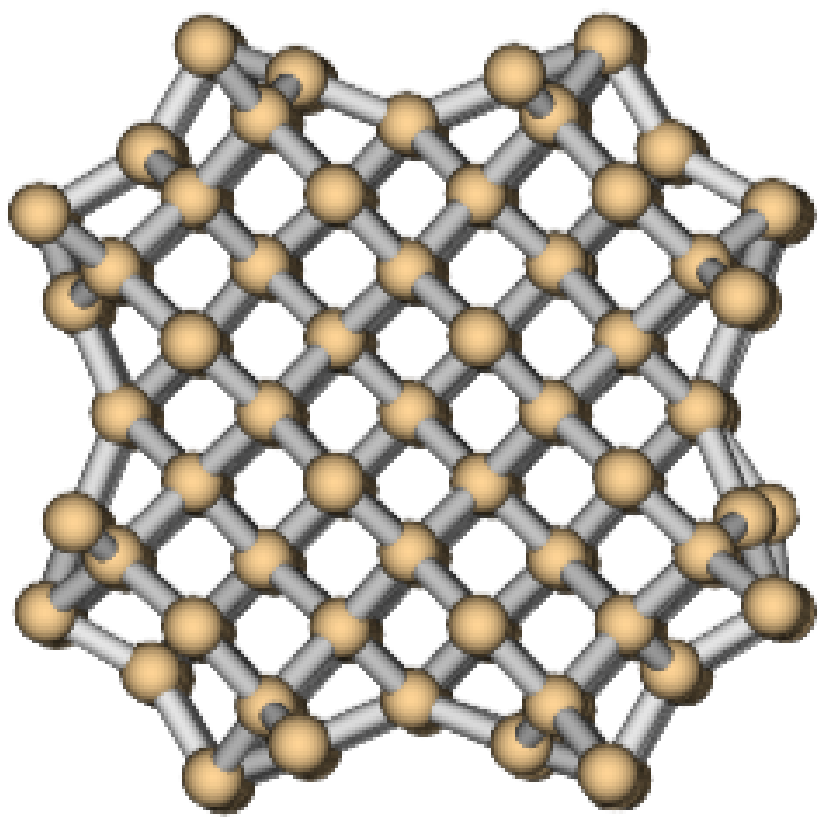}
\includegraphics[width=.26\textwidth]{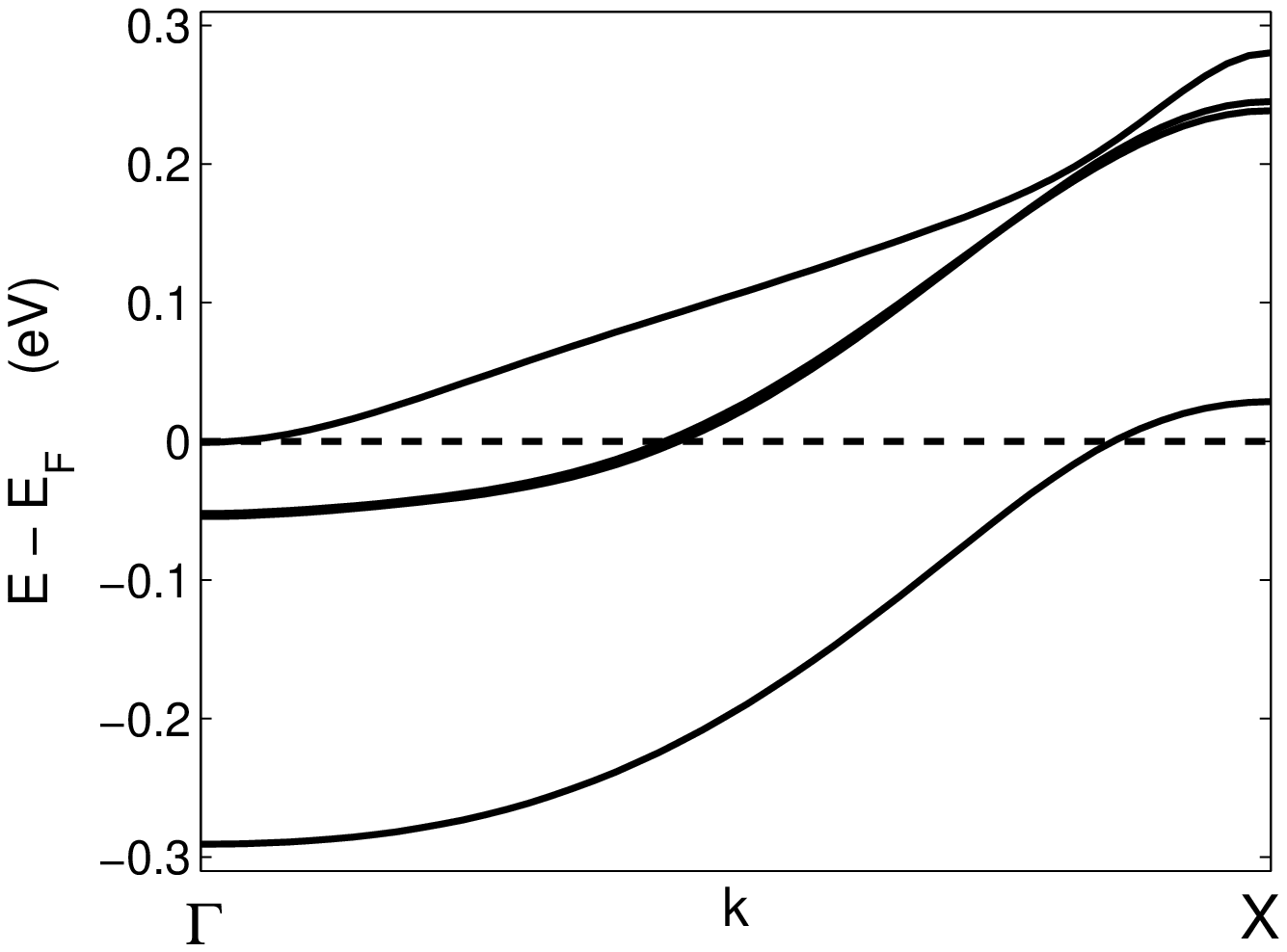}

	\caption{(Color online). Top: Side-view of the un-passivated SiNW containing 5 unit cells, separated by the red dashed lines. Bottom left: End view of the un-passivated SiNW. Bottom right: Band structure around the Fermi energy (dashed line). Four bands are crossing the Fermi energy, two being degenerate (with $E=-0.05\,$eV at the $\Gamma-$point).}
\label{wire-fig}
\end{figure}

The first structures we consider are un-passivated, surface reconstructed SiNWs as illustrated in Fig. \ref{wire-fig} showing a side-view (top) and a cross-section view (bottom left) of a wire containing 5 unit cells, separated by the red dashed lines in the upper panel. This structure was recently studied by Rurali \textit{et al.}\cite{NicolasPRL} using DFT calculations. There are 57 atoms in one unit cell, the diameter is $\sim\,$1.5 nm, and the length of the unit cell in the growth direction is $0.55\,$nm. Notice that the structure at the surface differs significantly from the bulk of the wire. Ref. \onlinecite{NicolasPRL} showed that this particular surface structure makes the wire metallic, which is evident in Fig. \ref{wire-fig} (bottom right) showing the band structure around the Fermi energy (marked by the dashed line). Four bands are crossing the Fermi level (two being degenerate).

To further investigate the metallic surface states and to compare the Kubo and the GF method we add random on-site noise to either the surface atoms only or to the bulk atoms only. The relevant diagonal elements in the Hamiltonian are thus changed according to
\begin{eqnarray}
\Hb_{ii}\rightarrow \Hb_{ii} + \delta_i,
\end{eqnarray}
where $\delta_i$ takes values with equal probability in the interval $[-\Delta\varepsilon/2;\Delta\varepsilon/2]$ with $\Delta\varepsilon$ being the disorder strength. This Anderson model for electronic disorder is simple and widely applied, very recently Zhong \textit{et al.}\cite{ZhongNanoLett2006} used it to model surface disorder in shell-doped nanowires. Whether this simple model adequately describes physical defects merits a separate discussion, given at the end of section \ref{squarewire}.

\begin{figure}[htb]
\includegraphics[width=.48\textwidth]{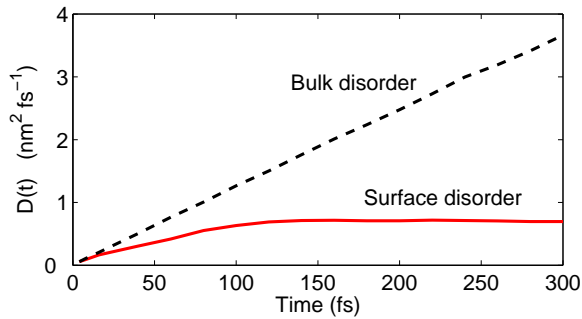}
\caption{(Color online). Time dependent diffusion coefficient at $E=E_F$. The bulk disorder has little effect and the transport is ballistic. The edge disordered system shows diffusive behaviour with a constant diffusion coefficient for $t>150\,$fs.}
\label{diffusion_bulk_surface}
\end{figure}

Fig. \ref{diffusion_bulk_surface} shows $\mathcal{D}(E_F,t)$ revealing that the bulk disorder has a very small effect on the transport properties yielding an almost linearly increasing diffusion coefficient, a characteristic feature of  ballistic transport. The surface disorder, on the other hand, leads to diffusive transport with $\mathcal{D}(E,t)$ almost constant for $t>150\,$fs. The pronounced differences of the surface- and bulk-disordered systems show that the conduction around the Fermi energy almost entirely takes place along the surface atoms, in qualitative agreement with the conclusions drawn in Ref. \onlinecite{NicolasPRL}.

\begin{figure}[htb]
\includegraphics[width=.44\textwidth]{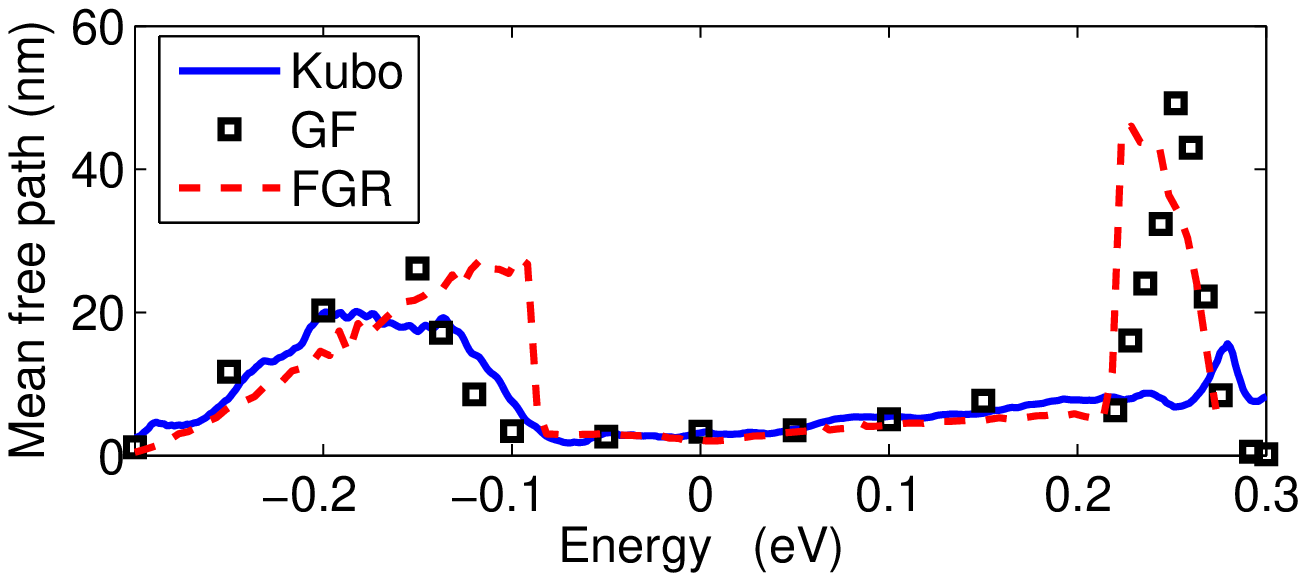}
\includegraphics[width=.44\textwidth]{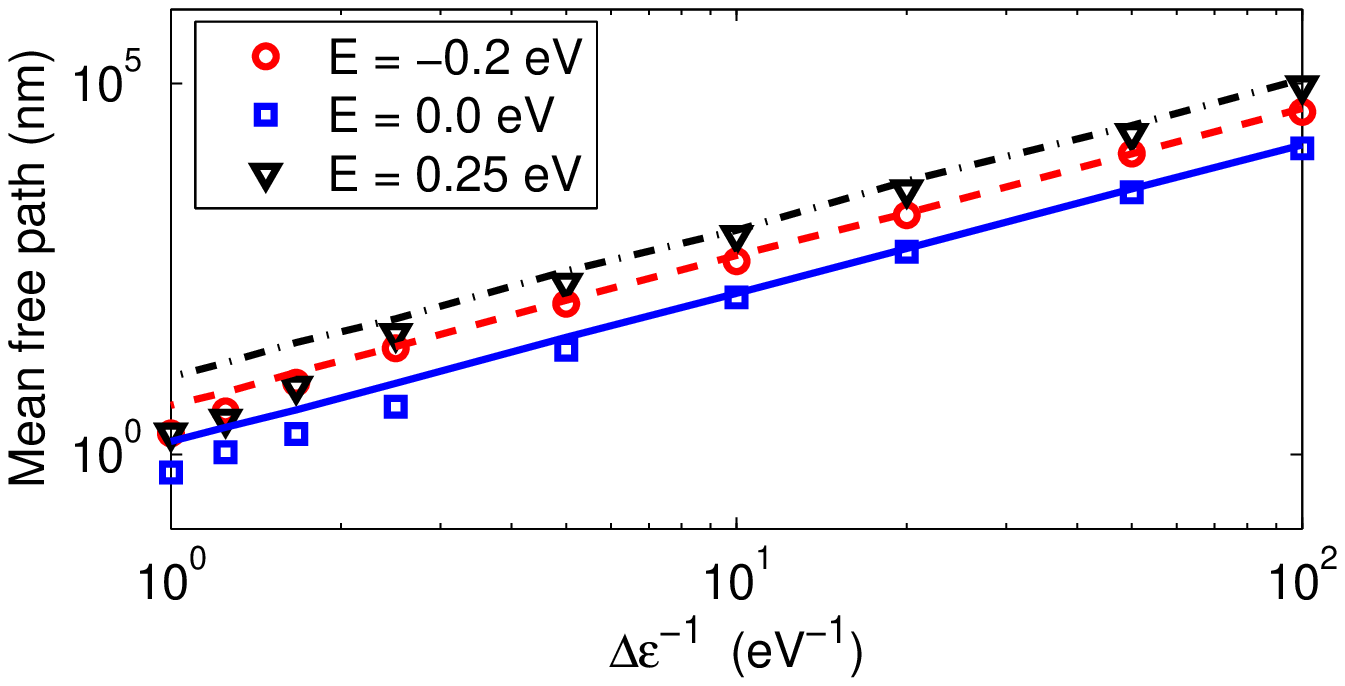}
\caption{(Color online). Top: Mean free path, $l_e$, vs. energy for $\Delta\varepsilon=0.4\,$eV. The solid lines are obtained with the Kubo method, the GF results are marked by squares while the dashed lines shows the FGR results. GF results are average values for 200 different samples while the Kubo results are mean values of 10 different samples. Bottom: Scaling of $l_e$ vs. $1/\Delta\varepsilon$ shown on a logarithmic scale. Circles, squares and triangles are calculated with GF while the lines are obtained using FGR at the same energies. }
\label{mfp_metallic}
\end{figure}

Fig. \ref{mfp_metallic} (upper panel) shows the MFP, $l_e(E)$, vs. energy calculated using both the Kubo method (solid blue) and the recursive GF approach (black squares). The dashed red line is an analytical estimate obtained using FGR. The disorder strength is $\Delta\varepsilon = 0.4\,$eV and only the surface  atoms are perturbed. The Kubo and FGR results are obtained with a fine energy resolution whereas the GF results are calculated only for relatively few, discrete energies, illustrating the advantages in the Kubo method in calculating properties at many energies in one calculation.

Generally all three methods agree qualitatively and in the interval $-0.1\lesssim E\lesssim 0.2$ the results are even quantitatively consistent. In this energy range several bands exist, see Fig. \ref{wire-fig} (bottom right) and thus there are more possible back scattering processes giving a larger scattering rate and thus a shorter MFP. Notice, that lowering the mean free path does not necessarily mean reducing the conductance, because we also have more charge carrying states.

Around the band edges at $E=-0.09\,$eV the MFP calculated with FGR drops sharply whereas the values obtained with the GF and Kubo methods drop more slowly. This difference is caused by the relatively large disorder strength, $\Delta\varepsilon=0.4\,$eV, which broadens the DOS and smears out the sharp features. For smaller disorder strengths the GF results resemble the FGR values more, which is illustrated in the lower panel in Fig. \ref{mfp_metallic}. The figure shows, on a logarithmic scale, the MFP vs. inverse disorder strength, $1/\Delta\varepsilon$, at three different energies. The points are obtained with the GF method, while the lines are obtained using FGR. For weak disorder, the GF results scales as $l_e\propto(1/\Delta\varepsilon)^2$ in accordance with FGR, whereas the GF results for strong disordered systems deviate from the $(1/\Delta\varepsilon)^2$ dependence. In this regime the first order perturbation applied in FGR does not fully suffices and reliable results must be obtained with more elaborate approaches such as the GF method.

Around $0.24\,$eV, the Kubo method fails to find the pronounced peak obtained with both the GF method and with FGR. This difference is probably caused by two effects. The first reason is again a broadened DOS, since the GF results show a similar deviation from the FGR values as seen around $E=-0.09\,$eV. The second and more important reason to the differences is due to the numerical inaccuracies in calculating the density of states in the Kubo method. The inaccuracies are especially important around van Hove singularities at the sub-band edges. $E=0.24\,$eV marks the band edge for the two degenerate bands, and due to the finiteness of the system considered in the Kubo calculation, the van Hove peaks in the DOS will unavoidably have decaying tails at larger energies. This implies that the calculated density of states will be too large causing  $\mathcal{D}(E,t)$  and thus $l_e$ to be correspondingly smaller. The energy separation between the two sub-band edges around $E=0.24\,$eV is only $\sim0.05\,$eV, which is less than $0.3\%$ of the total bandwidth, $W\approx20\,$eV. This makes it numerically difficult to resolve the detailed features with the continued fraction technique used by the Kubo method.


\subsection{Passivated wires} \label{squarewire}
The surface reconstructed SiNWs are both physically and technologically exciting, but probably also very fragile objects, since small changes in the surface such as defects or adatoms presumably will change the performance drastically. Moreover, the wires produced are most often surface passivated by either SiO$_x$ or hydrogen, and the focus in this section will be on such wires. The passivated wires are semiconducting, often with a direct band gap that increases for small diameters \cite{DDDMa, Lake2005, ZhaoPRL2004}. The simplest way to model surface passivated wires is to add hydrogen atoms to the surface such that all the Si dangling bonds are passivated. Such wires resemble qualitatively those reported by Ma \textit{et al.} \cite{DDDMa}.

\begin{figure}[htb]
\begin{minipage}[l]{.20\textwidth}
\begin{center}
    \includegraphics[width=.95\textwidth]{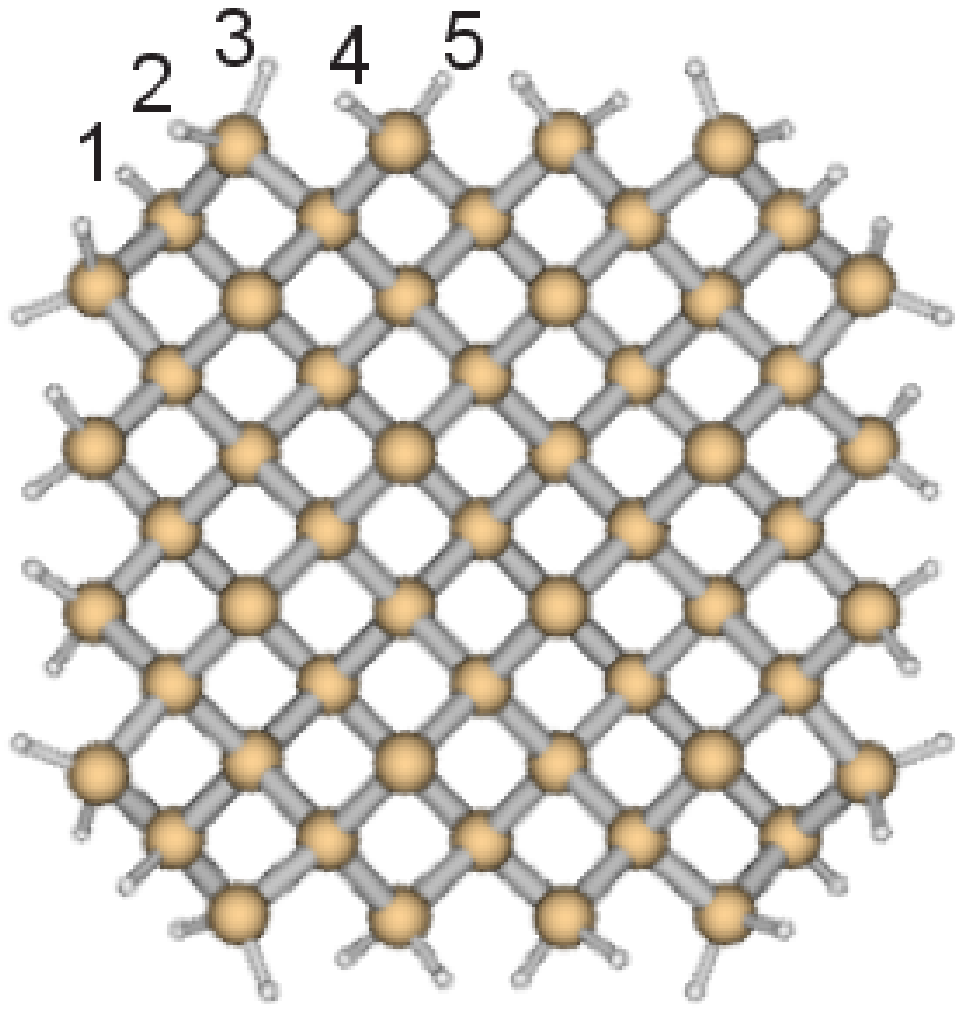}
\end{center}
\end{minipage}
\begin{minipage}[l]{.25\textwidth}
\begin{center}
    \includegraphics[angle=00, width=\textwidth]{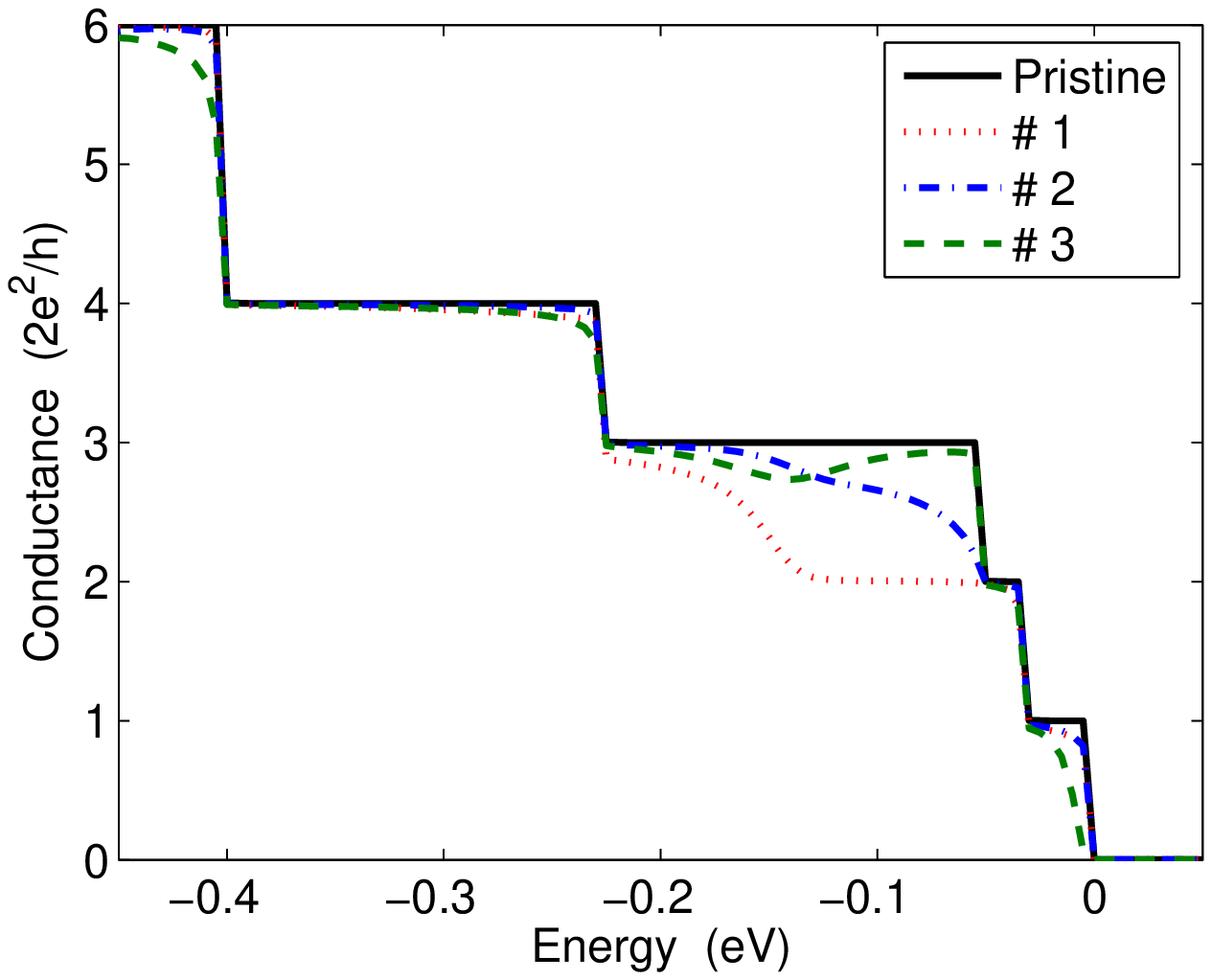}
\end{center}
\end{minipage}
\caption{(Color online). Left: Cross section of the H passivated wire. Right: Energy dependent conductance of a pristine wire and of wires containing a single vacancy of number 1-3.}
\label{wire-sq-fig}
\end{figure}

Figure \ref{wire-sq-fig} shows the cross section of the wire (left). The unit cell contains 57 Si atoms and 36 H atoms labeled with a number from 1 to 36 as indicated in Fig. \ref{wire-sq-fig} (left). The band gap is found to be $2.84\,$eV.\cite{bands}

To investigate the influence of surface defects we introduce hydrogen vacancies. The vacancies are labeled correspondingly to the removed hydrogen atoms. Note, that due to symmetry there are only five topologically different vacancies. The conductances of wires with only a single vacancy, corresponding to one of the H atoms 1-3 being removed, are shown in Fig. \ref{wire-sq-fig} (right) for energies close to the valence band edge ($E=0\,$eV). Clearly, vacancy 1 and 2 scatter the most, the reason probably being that the wave function (for the pristine wire) has a pronounced larger weight on these H atoms. Notice, for energies $-0.15<E<-0.05\,$eV, one channel is almost completely closed by vacancy 1. The vacancies 3-5 give almost the same conductances.

\begin{figure}[htb]
\includegraphics[width=.45\textwidth]{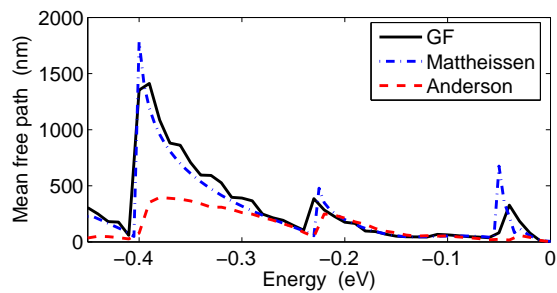}
\caption{(Color online). Mean free path vs. energy for a concentration of vacancies corresponding to $\langle d_H\rangle=5.6\,$nm. Solid black line corresponds to wires containing all possible vacancies. Dashed-dotted blue is obtained using Matthiessen's rule for single scattering events and the dashed red curve corresponds to a wire with Anderson disorder ($\Delta\varepsilon=1.3\,$eV).}
\label{mfp_vs_E_passivated}
\end{figure}

We model a wire with randomly missing hydrogen atoms by performing a \textsc{Siesta} calculation for each possible vacancy position (one of the 36 H atoms is removed) and adding pieces from the different calculations. We measure the vacancy concentration by the average distance, $\langle d_H\rangle$, in the wire direction between two vacancies. Each unit cell can only contain one vacancy thus setting a lower limit for $\langle d_H\rangle$ at the unit cell length, $a=0.56\,$nm. The MFP for $\langle d_H\rangle=5.6\,$nm calculated with the GF method is shown in Fig. \ref{mfp_vs_E_passivated} (solid) revealing a strong energy dependence. In the interval $-0.15<E<-0.05\,$eV, we find $l_e\sim50\,$nm, while for energies around $-0.35\,$eV the MFP in on the order of $1\,\mu$m. Comparing with estimated phonon scattering MFPs of more than 500 nm,\cite{LuLieberPNAS2005} the application of the elastic scattering model applied in this work is justified for most of the energies. Moreover, at several energies the vacancy scattering might be the dominant even at room temperature. However, around the peak at $E=-0.4\,$eV where the calculated MFPs exceeds $1\,\mu$m, other scattering sources are likely to dominate.

Assuming that all bands at a given energy have the same reflection probability, $R_i(E) = (T_0(E)-T_i(E))/T_0(E)$, where $T_0$ is the total transmission of a pristine wire, and $T_i$ is the total transmission through a wire containing a single vacancy with number $i$ (shown in Fig. \ref{wire-sq-fig} (left)),
the MFP in a wire with only vacancies of type $i$, can be estimated as $l_e^{(i)}(E)= \langle d_H\rangle/R_i(E)$. The total MFP can be estimated using Matthiessen's rule, such that $\tilde{l}_e^{-1}=\sum_i (l_e^{(i)})^{-1}$, and the result is shown in Fig.     \ref{mfp_vs_E_passivated} as the  dash-dotted line. It is evident that the simple estimate, which ignores interference effects between successive scatterers, accurately reproduces the results found by sample averaging over vacancy configurations.

The Kubo method requires a Hamiltonian describing a wire that is longer than the largest mean free path in order to avoid boundary effects. As seen from Fig. \ref{mfp_vs_E_passivated} we therefore need a wire of length $L>1500\,$nm consisting of more than 3,000 unit cells and thus $N>8\cdot10^5$ orbitals. In our current implementation this causes computer memory problems, and we have not been able to obtain reliable results with the Kubo method for the vacancy scattering.

We next examine whether the vacancies effectively can be modelled by adding Anderson disorder. The calculated MFP for a surface disordered wire, i.e. with no vacancies but rather on-site disorder at all orbitals at the surface Si atoms, is shown in Fig. \ref{mfp_vs_E_passivated} (dashed red).

For a disorder strength $\Delta\varepsilon=1.3\,$eV the Anderson model fits the vacancy results at energies $E>-0.35\,$eV, although the peak around $E=-0.05\,$eV is much less pronounced. The small shift of the peaks around $E=-0.23\,$ eV and  $E=-0.05\,$eV is due to a broadened DOS in the Anderson disordered wires. For energies below $-0.35\,$eV, the Anderson model deviates significantly from the vacancy results. The pronounced peak in MFP is not found in the Anderson model which gives almost constant values up to a factor of three lower than the vacancy results. Besides a broadened DOS the differences probably arise because the Anderson model is a too simple model not capturing all the  physics. Since the Anderson disorder only reproduces the vacancy results at some energies we conclude that the effects of vacancies cannot accurately be modelled by the simple on-site disorder. Moreover, the value of the disorder strength, $\Delta\varepsilon$, has no clear connection to an actual vacancy concentration.

\begin{figure}[htb]
\includegraphics[width=.45\textwidth]{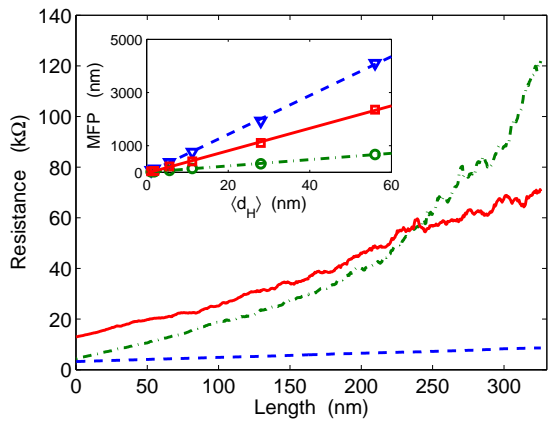}
\caption{(Color online). Length dependent resistance for at the energies $E=-0.3\,$eV (dashed blue), $E=-0.15\,$eV (dashed-dotted green) and $E=-0.03\,$eV (solid red). The average distance between vacancies is $\langle d_H\rangle=2.8\,$nm. The inset shows the scaling of $l_e$ vs. average distance, $\langle d_H\rangle$, between vacancies at the same three energies as above.}
\label{mfp_vs_dH}
\end{figure}

Fig. \ref{mfp_vs_dH} shows the length dependent resistance for three different energies at vacancy concentration corresponding to $\langle d_H\rangle=2.8\,$nm. For length $L<200\,$nm the resistance increases linearly at all energies, with the slope determining the MFP. The inset shows the scaling of $l_e$ vs. $\langle d_H\rangle$ at the same three energies as in the main figure. The points are calculated with the GF method and the lines are linear fits, clearly revealing a linear relationship between the MFP and the average inter-vacancy distance. For length $L>200\,$nm the resistance corresponding to $E=-0.15\,$eV (dash-dotted green) starts to increase exponentially thus entering the localization regime. The localization length at this energy is $\xi =110\,$nm, much lower than e.g. the MFP of $l_e\approx1\,\mu$m at slightly lower energies around $E=-0.35\,$eV.



\section{Discussion} \label{Conclusion}
In this paper we have studied electronic transport in SiNWs and calculated the influence of disorder on the mean free path. Our model is subject to a number of limitations and approximations. We apply a single-electron model and consider the linear response regime. Moreover, the minimal basis set may limit the accuracy. Also the spin orbit coupling is not included which is necessary to describe the details around the valence band edge in bulk Si. Compared to the experimentally realized SiNWs, the wires considered here are quite thin although comparable to the wires reported in Ref. \onlinecite{DDDMa}. The SiNW structures used in this paper have rounded, rather than perfectly square angles. It has been proposed in the literature~\cite{arias,yakobson,rurali_nt} that smooth angles would  be favored during the growth process with respect to the sharp angles that naturally arise from the square symmetry of the [100] cleavage plane. The topic has been discussed at some details elsewhere~\cite{rurali_nt} for surface reconstructed wires, concluding that at nanometric diameters no clear difference emerges. At the same time the electronic structure seems to be only marginally affected. Since all calculations in this work are performed on relaxed structures fully based on \textit{ab initio} calculations without any use of fitting parameters, we expect, in spite of all the limitations, to capture the correct \textit{trends} in the transport characteristics.

\subsection{Methods}
Two numerical methods were applied and compared to each other: A real-space Kubo method, and a recursive Green's function method. The two approaches each have their advantages: In calculating the MFP at many energies, the Kubo method is advantageous, since the diffusion is readily found at many energies in a single calculation, whereas the GF method requires a full calculation at each energy.  If the focus is on a few energies but many different disorders, the GF method is the preferred choice. For metallic systems, where one mainly is interested in the properties around the Fermi level, the GF method thus seems to be the method of choice - the parallel computation of many energies in the Kubo method is not needed.
The Kubo method seems more applicable to semiconducting systems, since it is physically possible to move the Fermi level with a gate voltage, thereby scanning several energies.

The Kubo method requires a Hamiltonian describing a wire that is longer than the largest MFP. For the weakly disordered wires with long MFPs we need wires of length $L\geq1\,\mu$m containing more than $10^5$ atoms resulting in very large matrices. In our current implementation this causes memory problems and the Kubo method failed to converge for the H-passivated SiNWs.
The GF method, on the other hand, involves only calculations with the small sub-cell Hamiltonians and it suffices to consider wires grown to a length $L\approx50\,$nm to get an accurate estimate of the initially linear resistance vs. length curve (cf. Fig. \ref{mfp_vs_dH}). Moreover, it proved to be numerically difficult to resolve the detailed features in the energy spectrum with the Kubo method, which led to erroneous results near band edges. The difficulties arise because the Kubo method does not take any semi-infinite periodic leads into account as in the GF method. There, the DOS is readily calculated to arbitrary accuracy from the surface Green's function $\Gb_L^{0}$ by using the periodic structure of the leads. However, the absence of periodic leads in the Kubo method can also be a great advantage since it allows to study non-periodic systems such as incommensurable multi-walled carbon nanotubes.\cite{TriozonPRB2004}

The growth procedure in the GF method involves inversions of the relatively small matrices describing the sub-cells. For thin wires as those considered in this work, with $N\sim250$ orbitals in each unit cell, the inversion step is not a critical issue. However, for thicker wires the number of orbitals increases quadratically, and due to an $\mathcal{O}(N^{3})$ scaling of the inversion step, the whole procedure scales as $\mathcal{O}(d^{\,6})$, with $d$ being the wire diameter. Contrary, the Kubo method scales linearly with the number of orbitals, and thus as $\mathcal{O}(d^{\,2})$.

For systems with a periodic structure, as the SiNWs considered here, we generally find the GF method to be the preferred choice, given that a rough energy resolution is sufficient. Thicker wires with relatively short MFPs favor the use of the Kubo approach.

\subsection{Results}
In un-passivated, surface reconstructed wires, Anderson disorder was added to the surface atoms affecting the transport properties significantly. Disorder in the bulk had, on the other hand, no significant influence.

In hydrogen passivated wires surface disorder was introduced by randomly removed hydrogen atoms. We find that it suffices to consider single scattering events and adding the individually calculated MFPs to an effective MFP by applying Matthiessen's rule. However, it is not obvious that a similar rule applies for the localization lengths thus emphasizing the importance of considering systems containing many differently placed vacancies as opposed to only considering single scattering events. 
It was further shown that an attempt to model the vacancies with an effective Anderson disorder gave satisfactory results only in a limited energy interval. We therefore conclude that the Anderson model is a too simplistic description of the real vacancies.

The MFP was shown to be strongly energy dependent, and for relatively long wires the resistance can be changed by orders of magnitude within a 0.1 eV shift of the Fermi energy thus causing a transition from the diffusive (Ohmic) regime to the localization regime. The strong energy dependence might be utilized in sensor applications where the presence of a single virus acts as a local gate shifting the Fermi energy.\cite{LieberMatToday2005} This could possibly cause a transition from the Ohmic to the localization regime thus changing the resistance of the wire dramatically. However, more careful work has to be done before firm conclusions can be stated.

For relatively strong disordered wires, the MFP is well below 500 nm for a large energy range. Compared to estimated phonon scattering MFPs of more than 500 nm\cite{LuLieberPNAS2005}, the elastic scattering model applied in this work is justified. Comparing the results obtained in this work with the estimated long phonon MFP we suggest that impurity and defect scattering could be the dominant scattering source even at room temperature.

\begin{acknowledgements}
We thank Profs. H. Smith and N. Lorente for discussions, and the Danish Center for Scientific Computing (DCSC) for providing
computer resources. T.M. thanks the Oticon Foundation for financial support. R.R. acknowledge financial support of the Ministerio de Educación y Ciencia through the Juan de la Cierva programme.
\end{acknowledgements}


\begin{thebibliography}{47}
\expandafter\ifx\csname natexlab\endcsname\relax\def\natexlab#1{#1}\fi
\expandafter\ifx\csname bibnamefont\endcsname\relax
  \def\bibnamefont#1{#1}\fi
\expandafter\ifx\csname bibfnamefont\endcsname\relax
  \def\bibfnamefont#1{#1}\fi
\expandafter\ifx\csname citenamefont\endcsname\relax
  \def\citenamefont#1{#1}\fi
\expandafter\ifx\csname url\endcsname\relax
  \def\url#1{\texttt{#1}}\fi
\expandafter\ifx\csname urlprefix\endcsname\relax\def\urlprefix{URL }\fi
\providecommand{\bibinfo}[2]{#2}
\providecommand{\eprint}[2][]{\url{#2}}

\bibitem[{\citenamefont{Cui et~al.}(2003)\citenamefont{Cui, Wang, Wang, and
  Lieber}}]{CuiLieberNanoLett2003}
\bibinfo{author}{\bibfnamefont{Y.}~\bibnamefont{Cui}},
  \bibinfo{author}{\bibfnamefont{D.}~\bibnamefont{Wang}},
  \bibinfo{author}{\bibfnamefont{W.~U.} \bibnamefont{Wang}}, \bibnamefont{and}
  \bibinfo{author}{\bibfnamefont{C.~M.} \bibnamefont{Lieber}},
  \bibinfo{journal}{Nano Lett.} \textbf{\bibinfo{volume}{3}},
  \bibinfo{pages}{149} (\bibinfo{year}{2003}).

\bibitem[{\citenamefont{Cui and Lieber}(2001)}]{LieberScience2001}
\bibinfo{author}{\bibfnamefont{Y.}~\bibnamefont{Cui}} \bibnamefont{and}
  \bibinfo{author}{\bibfnamefont{C.~M.} \bibnamefont{Lieber}},
  \bibinfo{journal}{Science} \textbf{\bibinfo{volume}{291}},
  \bibinfo{pages}{851} (\bibinfo{year}{2001}).

\bibitem[{\citenamefont{Gudiksen et~al.}(2002)\citenamefont{Gudiksen, Lauhon,
  Wang, Smith, and Lieber}}]{GudiksenLieber2002}
\bibinfo{author}{\bibfnamefont{M.~S.} \bibnamefont{Gudiksen}},
  \bibinfo{author}{\bibfnamefont{L.~J.} \bibnamefont{Lauhon}},
  \bibinfo{author}{\bibfnamefont{J.}~\bibnamefont{Wang}},
  \bibinfo{author}{\bibfnamefont{D.~C.} \bibnamefont{Smith}}, \bibnamefont{and}
  \bibinfo{author}{\bibfnamefont{C.~M.} \bibnamefont{Lieber}},
  \bibinfo{journal}{Nature} \textbf{\bibinfo{volume}{415}},
  \bibinfo{pages}{617} (\bibinfo{year}{2002}).

\bibitem[{\citenamefont{Samuelson}(2003)}]{SamuelsonMatToday}
\bibinfo{author}{\bibfnamefont{L.}~\bibnamefont{Samuelson}},
  \bibinfo{journal}{Materials Today} \textbf{\bibinfo{volume}{6}},
  \bibinfo{pages}{22} (\bibinfo{year}{2003}).

\bibitem[{\citenamefont{Samuelson et~al.}(2004)\citenamefont{Samuelson,
  Thelander, Björk, Borgström, Deppert, Dick, Hansen, Mårtensson, Panev,
  Persson et~al.}}]{SamuelsonPhysicaE}
\bibinfo{author}{\bibfnamefont{L.}~\bibnamefont{Samuelson}},
  \bibinfo{author}{\bibfnamefont{C.}~\bibnamefont{Thelander}},
  \bibinfo{author}{\bibfnamefont{M.~T.} \bibnamefont{Björk}},
  \bibinfo{author}{\bibfnamefont{M.}~\bibnamefont{Borgström}},
  \bibinfo{author}{\bibfnamefont{K.}~\bibnamefont{Deppert}},
  \bibinfo{author}{\bibfnamefont{K.}~\bibnamefont{Dick}},
  \bibinfo{author}{\bibfnamefont{A.}~\bibnamefont{Hansen}},
  \bibinfo{author}{\bibfnamefont{T.}~\bibnamefont{Mårtensson}},
  \bibinfo{author}{\bibfnamefont{N.}~\bibnamefont{Panev}},
  \bibinfo{author}{\bibfnamefont{A.}~\bibnamefont{Persson}},
  \bibnamefont{et~al.}, \bibinfo{journal}{Physica E}
  \textbf{\bibinfo{volume}{25}}, \bibinfo{pages}{313} (\bibinfo{year}{2004}).

\bibitem[{\citenamefont{Wu et~al.}(2004{\natexlab{a}})\citenamefont{Wu, Yang,
  Lu, and Lieber}}]{WuLieberNature2004}
\bibinfo{author}{\bibfnamefont{Y.}~\bibnamefont{Wu}},
  \bibinfo{author}{\bibfnamefont{C.}~\bibnamefont{Yang}},
  \bibinfo{author}{\bibfnamefont{W.}~\bibnamefont{Lu}}, \bibnamefont{and}
  \bibinfo{author}{\bibfnamefont{C.~M.} \bibnamefont{Lieber}},
  \bibinfo{journal}{Nature} \textbf{\bibinfo{volume}{430}}, \bibinfo{pages}{61}
  (\bibinfo{year}{2004}{\natexlab{a}}).

\bibitem[{\citenamefont{Wu et~al.}(2004{\natexlab{b}})\citenamefont{Wu, Cui,
  Huynh, Barrelet, Bell, and Lieber}}]{WuLieberNanoLett2004}
\bibinfo{author}{\bibfnamefont{Y.}~\bibnamefont{Wu}},
  \bibinfo{author}{\bibfnamefont{Y.}~\bibnamefont{Cui}},
  \bibinfo{author}{\bibfnamefont{L.}~\bibnamefont{Huynh}},
  \bibinfo{author}{\bibfnamefont{C.}~\bibnamefont{Barrelet}},
  \bibinfo{author}{\bibfnamefont{D.}~\bibnamefont{Bell}}, \bibnamefont{and}
  \bibinfo{author}{\bibfnamefont{C.}~\bibnamefont{Lieber}},
  \bibinfo{journal}{Nano Lett.} \textbf{\bibinfo{volume}{4}},
  \bibinfo{pages}{433} (\bibinfo{year}{2004}{\natexlab{b}}).

\bibitem[{\citenamefont{Patolsky and Lieber}(2005)}]{LieberMatToday2005}
\bibinfo{author}{\bibfnamefont{F.}~\bibnamefont{Patolsky}} \bibnamefont{and}
  \bibinfo{author}{\bibfnamefont{C.~M.} \bibnamefont{Lieber}},
  \bibinfo{journal}{Materials Today} \textbf{\bibinfo{volume}{8}},
  \bibinfo{pages}{20} (\bibinfo{year}{2005}).

\bibitem[{\citenamefont{Ma et~al.}(2003)\citenamefont{Ma, Lee, Au, Tong, and
  Lee}}]{DDDMa}
\bibinfo{author}{\bibfnamefont{D.~D.~D.} \bibnamefont{Ma}},
  \bibinfo{author}{\bibfnamefont{C.~S.} \bibnamefont{Lee}},
  \bibinfo{author}{\bibfnamefont{F.~K.} \bibnamefont{Au}},
  \bibinfo{author}{\bibfnamefont{S.~T.} \bibnamefont{Tong}}, \bibnamefont{and}
  \bibinfo{author}{\bibfnamefont{S.~T.} \bibnamefont{Lee}},
  \bibinfo{journal}{Science} \textbf{\bibinfo{volume}{299}},
  \bibinfo{pages}{1874} (\bibinfo{year}{2003}).

\bibitem[{\citenamefont{Holmes et~al.}(2000)\citenamefont{Holmes, Johnston,
  Doty, and Korgel}}]{HolmesScience2000}
\bibinfo{author}{\bibfnamefont{J.~D.} \bibnamefont{Holmes}},
  \bibinfo{author}{\bibfnamefont{K.}~\bibnamefont{Johnston}},
  \bibinfo{author}{\bibfnamefont{R.~C.} \bibnamefont{Doty}}, \bibnamefont{and}
  \bibinfo{author}{\bibfnamefont{B.~A.} \bibnamefont{Korgel}},
  \bibinfo{journal}{Science} \textbf{\bibinfo{volume}{287}},
  \bibinfo{pages}{1471} (\bibinfo{year}{2000}).

\bibitem[{\citenamefont{Cui et~al.}(2001)\citenamefont{Cui, Lauhon, Gudiksen,
  Wang, and Lieber}}]{LieberAPL2001}
\bibinfo{author}{\bibfnamefont{Y.}~\bibnamefont{Cui}},
  \bibinfo{author}{\bibfnamefont{L.~J.} \bibnamefont{Lauhon}},
  \bibinfo{author}{\bibfnamefont{M.~S.} \bibnamefont{Gudiksen}},
  \bibinfo{author}{\bibfnamefont{J.}~\bibnamefont{Wang}}, \bibnamefont{and}
  \bibinfo{author}{\bibfnamefont{C.~M.} \bibnamefont{Lieber}},
  \bibinfo{journal}{Appl. Phys. Lett.} \textbf{\bibinfo{volume}{78}},
  \bibinfo{pages}{2214} (\bibinfo{year}{2001}).

\bibitem[{\citenamefont{Rurali and Lorente}(2005{\natexlab{a}})}]{NicolasPRL}
\bibinfo{author}{\bibfnamefont{R.}~\bibnamefont{Rurali}} \bibnamefont{and}
  \bibinfo{author}{\bibfnamefont{N.}~\bibnamefont{Lorente}},
  \bibinfo{journal}{Phys. Rev. Lett.} \textbf{\bibinfo{volume}{94}},
  \bibinfo{pages}{026805} (\bibinfo{year}{2005}{\natexlab{a}}).

\bibitem[{\citenamefont{Singh et~al.}(2005)\citenamefont{Singh, Kumar, Note,
  and Kawazoe}}]{SinghNanoLett2005}
\bibinfo{author}{\bibfnamefont{A.~K.} \bibnamefont{Singh}},
  \bibinfo{author}{\bibfnamefont{V.}~\bibnamefont{Kumar}},
  \bibinfo{author}{\bibfnamefont{R.}~\bibnamefont{Note}}, \bibnamefont{and}
  \bibinfo{author}{\bibfnamefont{Y.}~\bibnamefont{Kawazoe}},
  \bibinfo{journal}{Nano Lett.} \textbf{\bibinfo{volume}{5}},
  \bibinfo{pages}{2302} (\bibinfo{year}{2005}).

\bibitem[{\citenamefont{Fernandez-Serra
  et~al.}(2006)\citenamefont{Fernandez-Serra, Adessi, and Blase}}]{BlasePRL}
\bibinfo{author}{\bibfnamefont{M.V.}~\bibnamefont{Fernandez-Serra}},
  \bibinfo{author}{\bibfnamefont{C.}~\bibnamefont{Adessi}}, \bibnamefont{and}
  \bibinfo{author}{\bibfnamefont{X.}~\bibnamefont{Blase}},
  \bibinfo{journal}{Phys. Rev. Lett.} \textbf{\bibinfo{volume}{96}},
  \bibinfo{pages}{166805} (\bibinfo{year}{2006}).

\bibitem[{\citenamefont{Lu et~al.}(2005)\citenamefont{Lu, Xiang, Timko, Wu, and
  Lieber}}]{LuLieberPNAS2005}
\bibinfo{author}{\bibfnamefont{W.}~\bibnamefont{Lu}},
  \bibinfo{author}{\bibfnamefont{J.}~\bibnamefont{Xiang}},
  \bibinfo{author}{\bibfnamefont{B.~P.} \bibnamefont{Timko}},
  \bibinfo{author}{\bibfnamefont{Y.}~\bibnamefont{Wu}}, \bibnamefont{and}
  \bibinfo{author}{\bibfnamefont{C.~M.} \bibnamefont{Lieber}},
  \bibinfo{journal}{Proc. Natl. Acad. Sci. USA} \textbf{\bibinfo{volume}{102}},
  \bibinfo{pages}{10046} (\bibinfo{year}{2005}).

\bibitem[{\citenamefont{Das and Mizel}(2005)}]{Das2005}
\bibinfo{author}{\bibfnamefont{K.~K.} \bibnamefont{Das}} \bibnamefont{and}
  \bibinfo{author}{\bibfnamefont{A.}~\bibnamefont{Mizel}}, \bibinfo{journal}{J.
  Phys.: Condens. Matter} \textbf{\bibinfo{volume}{17}}, \bibinfo{pages}{6675}
  (\bibinfo{year}{2005}).

\bibitem[{\citenamefont{Sundaram and Mizel}(2004)}]{Sundaram2004}
\bibinfo{author}{\bibfnamefont{V.~S.} \bibnamefont{Sundaram}} \bibnamefont{and}
  \bibinfo{author}{\bibfnamefont{A.}~\bibnamefont{Mizel}}, \bibinfo{journal}{J.
  Phys.: Condens. Matter} \textbf{\bibinfo{volume}{16}},
  \bibinfo{pages}{4697–4709} (\bibinfo{year}{2004}).

\bibitem[{\citenamefont{Zheng et~al.}(2005)\citenamefont{Zheng, Riva, Lake,
  Alam, Boykin, and Klimeck}}]{Lake2005}
\bibinfo{author}{\bibfnamefont{Y.}~\bibnamefont{Zheng}},
  \bibinfo{author}{\bibfnamefont{C.}~\bibnamefont{Riva}},
  \bibinfo{author}{\bibfnamefont{R.}~\bibnamefont{Lake}},
  \bibinfo{author}{\bibfnamefont{K.}~\bibnamefont{Alam}},
  \bibinfo{author}{\bibfnamefont{T.~B.} \bibnamefont{Boykin}},
  \bibnamefont{and} \bibinfo{author}{\bibfnamefont{G.}~\bibnamefont{Klimeck}},
  \bibinfo{journal}{IEEE Trans. Electron Devices}
  \textbf{\bibinfo{volume}{52}}, \bibinfo{pages}{1097} (\bibinfo{year}{2005}).

\bibitem[{\citenamefont{Triozon et~al.}(2004)\citenamefont{Triozon, Roche,
  Rubio, and Mayou}}]{TriozonPRB2004}
\bibinfo{author}{\bibfnamefont{F.}~\bibnamefont{Triozon}},
  \bibinfo{author}{\bibfnamefont{S.}~\bibnamefont{Roche}},
  \bibinfo{author}{\bibfnamefont{A.}~\bibnamefont{Rubio}}, \bibnamefont{and}
  \bibinfo{author}{\bibfnamefont{D.}~\bibnamefont{Mayou}},
  \bibinfo{journal}{Phys. Rev. B} \textbf{\bibinfo{volume}{69}},
  \bibinfo{pages}{121410(R)} (\bibinfo{year}{2004}).

\bibitem[{\citenamefont{Roche}(1999)}]{RochePRB99}
\bibinfo{author}{\bibfnamefont{S.}~\bibnamefont{Roche}},
  \bibinfo{journal}{Phys. Rev. B} \textbf{\bibinfo{volume}{59}},
  \bibinfo{pages}{2284} (\bibinfo{year}{1999}).

\bibitem[{\citenamefont{Roche and Mayou}(1997)}]{RocheMayouPRL97}
\bibinfo{author}{\bibfnamefont{S.}~\bibnamefont{Roche}} \bibnamefont{and}
  \bibinfo{author}{\bibfnamefont{D.}~\bibnamefont{Mayou}},
  \bibinfo{journal}{Phys. Rev. Lett.} \textbf{\bibinfo{volume}{79}},
  \bibinfo{pages}{2518} (\bibinfo{year}{1997}).

\bibitem[{\citenamefont{Mayou}(2000)}]{MayouPRL2000}
\bibinfo{author}{\bibfnamefont{D.}~\bibnamefont{Mayou}},
  \bibinfo{journal}{Phys. Rev. Lett} \textbf{\bibinfo{volume}{85}},
  \bibinfo{pages}{1290} (\bibinfo{year}{2000}).

\bibitem[{\citenamefont{Roche and Saito}(2001)}]{RochePRL2001}
\bibinfo{author}{\bibfnamefont{S.}~\bibnamefont{Roche}} \bibnamefont{and}
  \bibinfo{author}{\bibfnamefont{R.}~\bibnamefont{Saito}},
  \bibinfo{journal}{Phys. Rev. Lett} \textbf{\bibinfo{volume}{87}},
  \bibinfo{pages}{246803} (\bibinfo{year}{2001}).

\bibitem[{\citenamefont{Latil et~al.}(2004)\citenamefont{Latil, Roche, Mayou,
  and Charlier}}]{RochePRL2004}
\bibinfo{author}{\bibfnamefont{S.}~\bibnamefont{Latil}},
  \bibinfo{author}{\bibfnamefont{S.}~\bibnamefont{Roche}},
  \bibinfo{author}{\bibfnamefont{D.}~\bibnamefont{Mayou}}, \bibnamefont{and}
  \bibinfo{author}{\bibfnamefont{J.-C.} \bibnamefont{Charlier}},
  \bibinfo{journal}{Phys. Rev. Lett.} \textbf{\bibinfo{volume}{92}},
  \bibinfo{pages}{256805} (\bibinfo{year}{2004}).

\bibitem[{\citenamefont{Roche et~al.}((2005))\citenamefont{Roche, Jiang,
  Triozon, and Saito}}]{RochePRB2005}
\bibinfo{author}{\bibfnamefont{S.}~\bibnamefont{Roche}},
  \bibinfo{author}{\bibfnamefont{J.}~\bibnamefont{Jiang}},
  \bibinfo{author}{\bibfnamefont{F.}~\bibnamefont{Triozon}}, \bibnamefont{and}
  \bibinfo{author}{\bibfnamefont{R.}~\bibnamefont{Saito}},
  \bibinfo{journal}{Phys. Rev. B} \textbf{\bibinfo{volume}{72}},
  \bibinfo{pages}{113410} (\bibinfo{year}{(2005)}).

\bibitem[{\citenamefont{Roche et~al.}(2005)\citenamefont{Roche, Jiang, Triozon,
  and Saito}}]{RochePRL2005}
\bibinfo{author}{\bibfnamefont{S.}~\bibnamefont{Roche}},
  \bibinfo{author}{\bibfnamefont{J.}~\bibnamefont{Jiang}},
  \bibinfo{author}{\bibfnamefont{F.}~\bibnamefont{Triozon}}, \bibnamefont{and}
  \bibinfo{author}{\bibfnamefont{R.}~\bibnamefont{Saito}},
  \bibinfo{journal}{Phys. Rev. Lett} \textbf{\bibinfo{volume}{95}},
  \bibinfo{pages}{076803} (\bibinfo{year}{2005}).

\bibitem[{\citenamefont{Latil et~al.}(2005)\citenamefont{Latil, Roche, and
  Charlier}}]{RocheNanoLett2005}
\bibinfo{author}{\bibfnamefont{S.}~\bibnamefont{Latil}},
  \bibinfo{author}{\bibfnamefont{S.}~\bibnamefont{Roche}}, \bibnamefont{and}
  \bibinfo{author}{\bibfnamefont{J.-C.} \bibnamefont{Charlier}},
  \bibinfo{journal}{Nano Lett.} \textbf{\bibinfo{volume}{5}},
  \bibinfo{pages}{2216} (\bibinfo{year}{2005}).

\bibitem[{\citenamefont{Kubo}(1957)}]{Kubo1957}
\bibinfo{author}{\bibfnamefont{R.}~\bibnamefont{Kubo}}, \bibinfo{journal}{J.
  Phys. Soc. Jpn.} \textbf{\bibinfo{volume}{12}}, \bibinfo{pages}{570}
  (\bibinfo{year}{1957}).

\bibitem[{\citenamefont{Greenwood}(1958)}]{Greenwood}
\bibinfo{author}{\bibfnamefont{D.}~\bibnamefont{Greenwood}},
  \bibinfo{journal}{Proc. Phys. Soc. London} \textbf{\bibinfo{volume}{71}},
  \bibinfo{pages}{585} (\bibinfo{year}{1958}).

\bibitem[{\citenamefont{White and Todorov}(1998)}]{TodorovNature}
\bibinfo{author}{\bibfnamefont{C.~T.} \bibnamefont{White}} \bibnamefont{and}
  \bibinfo{author}{\bibfnamefont{T.~N.} \bibnamefont{Todorov}},
  \bibinfo{journal}{Nature} \textbf{\bibinfo{volume}{393}},
  \bibinfo{pages}{240} (\bibinfo{year}{1998}).

\bibitem[{\citenamefont{Triozon et~al.}(2002)\citenamefont{Triozon, Vidal,
  Mosseri, and Mayou}}]{TriozonPRB2002}
\bibinfo{author}{\bibfnamefont{F.}~\bibnamefont{Triozon}},
  \bibinfo{author}{\bibfnamefont{J.}~\bibnamefont{Vidal}},
  \bibinfo{author}{\bibfnamefont{R.}~\bibnamefont{Mosseri}}, \bibnamefont{and}
  \bibinfo{author}{\bibfnamefont{D.}~\bibnamefont{Mayou}},
  \bibinfo{journal}{Phys. Rev. B} \textbf{\bibinfo{volume}{65}},
  \bibinfo{pages}{220202(R)} (\bibinfo{year}{2002}).

\bibitem[{\citenamefont{Todorov}(1996)}]{Todorov}
\bibinfo{author}{\bibfnamefont{T.~N.} \bibnamefont{Todorov}},
  \bibinfo{journal}{Phys. Rev. B} \textbf{\bibinfo{volume}{54}},
  \bibinfo{pages}{5801} (\bibinfo{year}{1996}).

\bibitem[{\citenamefont{Lopez-Sancho et~al.}(1984)\citenamefont{Lopez-Sancho,
  Lopez-Sancho, and Rubio}}]{LoLoRu.84.}
\bibinfo{author}{\bibfnamefont{M.}~\bibnamefont{Lopez-Sancho}},
  \bibinfo{author}{\bibfnamefont{J.}~\bibnamefont{Lopez-Sancho}},
  \bibnamefont{and} \bibinfo{author}{\bibfnamefont{J.}~\bibnamefont{Rubio}},
  \bibinfo{journal}{J. Phys. F} \textbf{\bibinfo{volume}{14}},
  \bibinfo{pages}{1205} (\bibinfo{year}{1984}).

\bibitem[{\citenamefont{Datta}(1995)}]{Datta}
\bibinfo{author}{\bibfnamefont{S.}~\bibnamefont{Datta}},
  \emph{\bibinfo{title}{Electronic Transport in Mesoscopic Systems}}
  (\bibinfo{publisher}{Cambridge University Press}, \bibinfo{year}{1995}).

\bibitem[{\citenamefont{Kramer and MacKinnon}(1993)}]{MacKinnon}
\bibinfo{author}{\bibfnamefont{B.}~\bibnamefont{Kramer}} \bibnamefont{and}
  \bibinfo{author}{\bibfnamefont{A.}~\bibnamefont{MacKinnon}},
  \bibinfo{journal}{Reports on Progress in Physics}
  \textbf{\bibinfo{volume}{56}}, \bibinfo{pages}{1469} (\bibinfo{year}{1993}).

\bibitem[{\citenamefont{Soler et~al.}(2002)\citenamefont{Soler, Artacho, Gale,
  García, Junquera, Ordejón, and Sánchez-Portal}}]{siesta-ref}
\bibinfo{author}{\bibfnamefont{J.~M.} \bibnamefont{Soler}},
  \bibinfo{author}{\bibfnamefont{E.}~\bibnamefont{Artacho}},
  \bibinfo{author}{\bibfnamefont{J.~D.} \bibnamefont{Gale}},
  \bibinfo{author}{\bibfnamefont{A.}~\bibnamefont{García}},
  \bibinfo{author}{\bibfnamefont{J.}~\bibnamefont{Junquera}},
  \bibinfo{author}{\bibfnamefont{P.}~\bibnamefont{Ordejón}}, \bibnamefont{and}
  \bibinfo{author}{\bibfnamefont{D.}~\bibnamefont{Sánchez-Portal}},
  \bibinfo{journal}{J. Phys.: Condens. Matter} \textbf{\bibinfo{volume}{14}},
  \bibinfo{pages}{2745} (\bibinfo{year}{2002}).

\bibitem[{\citenamefont{Artacho et~al.}(1999)\citenamefont{Artacho,
  Sánchez-Portal, Ordejón, García, and Soler}}]{SiestaBasis}
\bibinfo{author}{\bibfnamefont{E.}~\bibnamefont{Artacho}},
  \bibinfo{author}{\bibfnamefont{D.}~\bibnamefont{Sánchez-Portal}},
  \bibinfo{author}{\bibfnamefont{P.}~\bibnamefont{Ordejón}},
  \bibinfo{author}{\bibfnamefont{A.}~\bibnamefont{García}}, \bibnamefont{and}
  \bibinfo{author}{\bibfnamefont{J.~M.} \bibnamefont{Soler}},
  \bibinfo{journal}{Phys. Stat. Sol. (b)} \textbf{\bibinfo{volume}{215}},
  \bibinfo{pages}{809} (\bibinfo{year}{1999}).

\bibitem[{\citenamefont{Troullier and Martins}(1991)}]{pseudo}
\bibinfo{author}{\bibfnamefont{N.}~\bibnamefont{Troullier}} \bibnamefont{and}
  \bibinfo{author}{\bibfnamefont{J.~L.} \bibnamefont{Martins}},
  \bibinfo{journal}{Phys. Rev. B} \textbf{\bibinfo{volume}{43}},
  \bibinfo{pages}{1993} (\bibinfo{year}{1991}).

\bibitem[{\citenamefont{Perdew et~al.}(1996)\citenamefont{Perdew, Burke, and
  Ernzerhof}}]{gga}
\bibinfo{author}{\bibfnamefont{J.~P.} \bibnamefont{Perdew}},
  \bibinfo{author}{\bibfnamefont{K.}~\bibnamefont{Burke}}, \bibnamefont{and}
  \bibinfo{author}{\bibfnamefont{M.}~\bibnamefont{Ernzerhof}},
  \bibinfo{journal}{Phys. Rev. Lett.} \textbf{\bibinfo{volume}{77}},
  \bibinfo{pages}{3865} (\bibinfo{year}{1996}).

\bibitem[{\citenamefont{Monkhorst and Pack}(1973)}]{monk-pack}
\bibinfo{author}{\bibfnamefont{H.~J.} \bibnamefont{Monkhorst}}
  \bibnamefont{and} \bibinfo{author}{\bibfnamefont{J.~D.} \bibnamefont{Pack}},
  \bibinfo{journal}{Phys. Rev. B} \textbf{\bibinfo{volume}{8}},
  \bibinfo{pages}{5747} (\bibinfo{year}{1973}).

\bibitem[{\citenamefont{Zhong and Stocks}(2006)}]{ZhongNanoLett2006}
\bibinfo{author}{\bibfnamefont{J.}~\bibnamefont{Zhong}} \bibnamefont{and}
  \bibinfo{author}{\bibfnamefont{G.~M.} \bibnamefont{Stocks}},
  \bibinfo{journal}{Nano Lett.} \textbf{\bibinfo{volume}{6}},
  \bibinfo{pages}{128} (\bibinfo{year}{2006}).

\bibitem[{\citenamefont{Zhao et~al.}(2004)\citenamefont{Zhao, Wei, Yang, and
  Chou}}]{ZhaoPRL2004}
\bibinfo{author}{\bibfnamefont{X.}~\bibnamefont{Zhao}},
  \bibinfo{author}{\bibfnamefont{C.~M.} \bibnamefont{Wei}},
  \bibinfo{author}{\bibfnamefont{L.}~\bibnamefont{Yang}}, \bibnamefont{and}
  \bibinfo{author}{\bibfnamefont{M.~Y.} \bibnamefont{Chou}},
  \bibinfo{journal}{Phys. Rev. Lett.} \textbf{\bibinfo{volume}{92}},
  \bibinfo{pages}{236805} (\bibinfo{year}{2004}).

\bibitem[{ban()}]{bands}
\bibinfo{note}{The calculated band gap is larger than found in Ref.
  \onlinecite{SinghNanoLett2006} for a similar structure using a plane wave DFT
  method.\cite{bands} The band gap reduces to $\sim$~1.66~eV when using a
  double-$\zeta$ polarized basis set, in qualitiative agreement with the
  plane-wave results of Ref.~\onlinecite{SinghNanoLett2006} (the residual
  difference must be attributed to the slight differences in the wire
  structures).}

\bibitem[{\citenamefont{Ismail-Beigi and Arias}(1998)}]{arias}
\bibinfo{author}{\bibfnamefont{S.}~\bibnamefont{Ismail-Beigi}}
  \bibnamefont{and} \bibinfo{author}{\bibfnamefont{T.}~\bibnamefont{Arias}},
  \bibinfo{journal}{Phys. Rev. B} \textbf{\bibinfo{volume}{57}},
  \bibinfo{pages}{11923} (\bibinfo{year}{1998}).

\bibitem[{\citenamefont{Zhao and Yakobson}(2003)}]{yakobson}
\bibinfo{author}{\bibfnamefont{Y.}~\bibnamefont{Zhao}} \bibnamefont{and}
  \bibinfo{author}{\bibfnamefont{B.~I.} \bibnamefont{Yakobson}},
  \bibinfo{journal}{Phys. Rev. Lett.} \textbf{\bibinfo{volume}{91}},
  \bibinfo{pages}{035501} (\bibinfo{year}{2003}).

\bibitem[{\citenamefont{Rurali and Lorente}(2005{\natexlab{b}})}]{rurali_nt}
\bibinfo{author}{\bibfnamefont{R.}~\bibnamefont{Rurali}} \bibnamefont{and}
  \bibinfo{author}{\bibfnamefont{N.}~\bibnamefont{Lorente}},
  \bibinfo{journal}{Nanotech.} \textbf{\bibinfo{volume}{16}},
  \bibinfo{pages}{S250} (\bibinfo{year}{2005}{\natexlab{b}}).

\bibitem[{\citenamefont{Singh et~al.}(2006)\citenamefont{Singh, Kumar, Note,
  and Kawazoe}}]{SinghNanoLett2006}
\bibinfo{author}{\bibfnamefont{A.~K.} \bibnamefont{Singh}},
  \bibinfo{author}{\bibfnamefont{V.}~\bibnamefont{Kumar}},
  \bibinfo{author}{\bibfnamefont{R.}~\bibnamefont{Note}}, \bibnamefont{and}
  \bibinfo{author}{\bibfnamefont{Y.}~\bibnamefont{Kawazoe}},
  \bibinfo{journal}{Nano Lett.} \textbf{\bibinfo{volume}{6}},
  \bibinfo{pages}{920} (\bibinfo{year}{2006}).

\end{thebibliography}

\end{document}